\documentclass[twoside,8pt]{article}
\usepackage{epsfig}

\newcommand{\be}{\begin{equation}}
\newcommand{\ee}{\end{equation}}
\newcommand{\bea}{\begin{eqnarray}}
\newcommand{\eea}{\end{eqnarray}}

\begin{document}

\title{The stability of relativistic stars and the role of the adiabatic index}

\author{Ch.C.  Moustakidis\\
Department of Theoretical Physics, Aristotle University of
Thessaloniki, \\ 54124 Thessaloniki, Greece }

\maketitle

\begin{abstract}
We study the stability of  three analytical solutions of the Einstein's field equations for spheres of fluid. These solutions are suitable to describe compact objects including  white dwarfs, neutron stars and supermassive stars and they have been extensively employed in the literature.  We re-examine the range of stability of the Tolman VII solution,  we focus on the stability of the Buchdahl solution which is under contradiction in the literature and   we examine the stability of the Nariai IV  solution.  We found that all the mentioned solutions are stable in an extensive range of the compactness parameter. We also concentrate  on the effect of the adiabatic index on the instability condition. We found that the critical adiabatic index, depends linearly on the ratio of central pressure over central energy density  $P_c/{\cal E}_c$, up to high values of the compactness. Finally, we examine the possibility to impose constraints, via the adiabatic index,  on realistic equations of state in order to ensure stable configurations of compact objects.

\vspace{0.3cm}

PACS number(s):
04.40.Dg, 04.20.Jb, 26.60.-c, 26.60.Kp  
\end{abstract}

\section{Introduction}
It is well known that  relativistic stars exhibit some special properties, related to their structure, in comparison  with the Newtonian ones~\cite{Chandrasekhar-64,Chandrasekhar-64b,Fowler-64,Fowler-66,Bardeen-66,Zeldovich-65,Tooper-65}. One of these properties,  is the condition for stable configurations. The stability of  relativistic stars is a longstanding problem~\cite{Weinberg-72,Harrison-65,Zeldovich-78,Shapiro-83,Glendenning-2000,Haensel-07,Friedman-2013}.
There are various methods to examine the stability of a relativistic star~\cite{Bardeen-66}. In the present work we will employ the variational method  which has been developed  by Chandrasekhar~\cite{Chandrasekhar-64,Chandrasekhar-64b}. In particular, we will examine the stability of three analytical solutions of the Einstein's field equation for a spherical relativistic fluid.

It is worth to pointing out that it is more natural to use a realistic equation of state of the fluid interior in order to solve the Einstein's field equations. It has been remarked  by Tolman~\cite{Tolman-39}, that by adopting  a mathematically instead of a  physically motivated method to solve the field equations may lead to not physically interesting solutions. However, this method has the advantage that by having an explicit solution it then becomes easier to examine the implied physics. We should remark  that the analytical solutions are {\it source}
of infinite number of equations of state (plausible or not). Consequently, they can be used extensively,   in order to introduce and to establish  some universal approximations.

It has been also remarked  by Tolman~\cite{Tolman-39} that the static character of an analytical solution is self sufficient  to assure that the solution describes a possible state of equilibrium for a fluid, but is not sufficient to tell us if the  state of equilibrium would be stable towards disturbances. In addition, the question of stability is important since an unstable solution can not be able to describe a physically permanent state and consequently is not of physical interest.

In the present work we will employ the Tolman VII~\cite{Tolman-39}, the Buchdahl~\cite{Buchdal-67} and the Nariai IV ~\cite{Nariai-50,Nariai-51,Nariai-99} solutions. These solutions are suitable to describe normal neutron stars and in general very compact objects. They have the specific property  that the derived  density and pressure of the fluid interior  vanish at the surface.
The three mentioned solutions  have been extensively employed in the literature for the study of very compact objects (including mainly neutron stars)~\cite{Lattimer-2001,Lattimer-05,Lattimer-05a,Postnikov-010,Raghoonundun-15,Moustakidis-016}. In particular, the Tolman VII solution is the most popular one, with applications in various problems related to very compact object (in contradiction with the  Tolman's statement~\cite{Tolman-39}  that this solution, due to the complicate dependence of the pressure on the distance $r$, is not a convenient one for physical consideration). The stability of  the Tolman VII solution has been examined by Negi~\cite{Negi-1999,Negi-2001}. It was found that the solution is stable for a large range of the compactness parameter  $\beta=GM/Rc^2$.
The Buchdahl solution has been also employed in the literature in related studies and is a subject of relevant textbooks and review papers~\cite{Schutz-85,Lattimer-2000,Lattimer-07b,Lattimer-10b}. However, there is a contradiction in the literature concerning the stability of the Buchdahl solution. Firstly,  Knutsen~\cite{Knutsen-88} found out  that this solution is unstable. Later on, Negi~\cite{Negi-2007} re-examined the instability of the this solution. He found that it remains stable for a large range of the compactness parameter. The Nariai IV solution is the  most complicated, compared to the previous ones and is used less in the literature. The stability of Narai IV solution, according to our knowledge, has never been examined in the past.
Moreover, and this is one of the motivations of the present work, we will investigate the possibility, using  physical reliable analytical solutions,  to introduce some general constraints related with the internal structure and the stability of relativistic objects.

The adiabatic index is  an important parameter which characterizes the stiffness of the equation of states  at a given density~\cite{Harrison-65,Haensel-07,Bludman-73a,Bludman-73b,Ipser-70}. The instability criterion of Chandrasekhar~\cite{Chandrasekhar-64,Chandrasekhar-64b}, which is an elegant interlay between the macroscopic attractive gravitation field and the microscopic repulsive nuclear forces,   strongly depends on the adiabatic index. In the present work we intend to probe the  possibility to impose constraints on the neutron star equation of state via  the instability condition of Chandrasekhar.

The article is organized as follows. In Section~2 we review briefly the Einstein's field equations while in  Section~ 3 we present  the Chandrasekhar instability criterion. Section~4 is dedicated to the adiabatic index and in Section~5 we present in detail  the analytical solutions.
The results are presented and discussed in Section~6 and Section~7 contains the concluding remarks of the study.

\section{The Einstein's field equations of a spherical fluid}
For a static spherical symmetric system, the metric can be written as follow~\cite{Shapiro-83,Glendenning-2000}
\begin{equation}
ds^2=e^{\nu(r)}dt^2-e^{\lambda(r)}dr^2-r^2\left(d\theta^2+\sin^2\theta d\phi^2\right).
\label{GRE-1}
\end{equation}
The  energy density distribution ${\cal E}(r)$ and the local pressure $P(r)$ are related to the  metric functions $\lambda(r)$ and $\nu(r)$ as follows~\cite{Shapiro-83,Glendenning-2000}
\begin{equation}
\frac{8\pi G}{c^4}{\cal E}(r)=\frac{1}{r^2}\left(1-e^{-\lambda(r)}\right)+ e^{-\lambda(r)}\frac{\lambda'(r)}{r},
\label{GRE-2}
\end{equation}
\begin{equation}
\frac{8\pi G}{c^4}P(r)=-\frac{1}{r^2}\left(1-e^{-\lambda(r)}\right)+ e^{-\lambda(r)}\frac{\nu'(r)}{r},
\label{GRE-3}
\end{equation}
\begin{equation}
P'(r)=-\frac{P(r)+{\cal E}(r)}{2}\nu'(r)
\label{GRE-4}
\end{equation}
where derivatives with respect to the radius are denoted by $'$. The mechanical equilibrium of the star matter is determined by the three equations
together with the equation of state ${\cal E}={\cal E}(P)$ of the fluid. The combination of Eqs.~(\ref{GRE-2}), (\ref{GRE-3}) and (\ref{GRE-4})  leads to the  well known
Tolman-Oppenheimer-Volkoff (TOV) equations~\cite{Shapiro-83,Glendenning-2000,Tolman-39,Oppenheimer-39}
\begin{equation}
\frac{dP(r)}{dr}=-\frac{G{\cal E}(r) M(r)}{c^2r^2}\left(1+\frac{P(r)}{{\cal E}(r)}\right)\left(1+\frac{4\pi P(r) r^3}{M(r)c^2}\right) \left(1-\frac{2GM(r)}{c^2r}\right)^{-1},
\label{TOV-1}
\end{equation}
\begin{equation}
\frac{dM(r)}{dr}=\frac{4\pi r^2}{c^2}{\cal E}(r).
\label{TOV-2}
\end{equation}
It is most natural to solve numerically  the TOV equation, by introducing an equation of state describing the relation between pressure and density expected to describe the fluid interior. The other possibility, is to try to find out analytical solutions of the TOV equation with the risk of obtaining solutions without physical interest~\cite{Tolman-39}.
Actually, there are hundreds of analytical solutions of TOV equations~\cite{Kramer-1980,Delgaty-1998,Lake-2003}. However, just few of them are of physical interest. Moreover, there are only three that satisfy the criteria that the pressure and energy density vanish on the surface of the star and also that  they both decrease monotonically with increasing radius. These three solutions are  the Tolman VII, the  Buchdahl and the Nariai IV  and  summarized in Section~5.

\section{The dynamical instability criterion of Chandrasekhar }
The dynamical stability may be assured by using the variational method. According to this method the sufficient condition for the dynamical stability of a mass is that the right-hand side of the equation~\cite{Chandrasekhar-64}
\begin{eqnarray}
\sigma^2\int_{0}^R e^{(3\lambda-\nu)/2}(P+ {\cal E}) r^2 \xi^2  dr&=&4\int_{0}^R e^{(\lambda+\nu)/2} r\frac{ dP}{dr}\xi^2 dr+
\int_{0}^R e^{(\lambda+3\nu)/2}\frac{\gamma P}{r^2}\left(\frac{d}{dr}(r^2e^{-\nu/2}\xi)  \right)^2 dr \nonumber \\
&-& \int_{0}^R e^{(\lambda+\nu)/2}\left(\frac{d P}{d r}  \right)^2\frac{r^2\xi^2}{P+ {\cal E}} dr+
\frac{8\pi G}{c^4}\int_0^R e^{(3\lambda+\nu)/2}P (P+{\cal E}) r^2\xi^2 dr
\label{cond-1}
\end{eqnarray}
vanishes for some chosen {\it trial function} $\xi(r)$ which satisfies the boundary conditions
\begin{equation}
\xi=0 \quad   {\rm at} \quad  r=0 \quad  {\rm and} \quad \delta P=0 \quad {\rm at} \quad r=R.
\label{cond-2}
\end{equation}
Obviously, there are  infinite numbers of  {\it trial functions} which satisfy the conditions~(\ref{cond-2}). The most well used are the following
\begin{equation}
\xi(r)=b_1r\left(1+a_1r^2+a_2r^4+a_3r^6\right)e^{\nu/2},
\label{xi-1}
\end{equation}
\begin{equation}
\xi(r)=r e^{\nu/4}
\label{xi-2}
\end{equation}
and also
\begin{equation}
\xi(r)=r e^{\nu/2}.
\label{xi-3}
\end{equation}
The function $\xi$ represents the amplitude of the lagrangian displacement from equilibrium, which has the form $\zeta=\xi(r)e^{-i\sigma t}$ where $\sigma$ represents the pulsation of the oscillations. In particular, when the squared pulsation $\sigma^2$ is negative the configuration is unstable since the displacement $\zeta$ grows exponentially in time. When the pulsation is positive $\zeta$ decreases also exponentially and the configuration is stable~\cite{Chandrasekhar-64,Merafina-89}. The stability condition~(\ref{cond-1})  expresses a minimal, and not merely an external principle~\cite{Chandrasekhar-64}. A comprehensive discussion of various methods for studying the normal modes of radial pulsation of General-Relativistic stellar models is presented by Bardeen {\it et al}.~\cite{Bardeen-66}. In reference~\cite{Bardeen-66} the advantage and disadvantage of the present method are also analyzed and discussed.

For an adiabatic perturbation, the adiabatic index $\gamma$, which appears in the stability formulae~(\ref{cond-1}) can be expressed in the form~\cite{Chandrasekhar-64,Merafina-89}
\begin{equation}
\gamma\equiv \frac{P+{\cal E}}{P}\left(\frac{\partial P}{\partial{\cal E}}\right)_S=\left(1+\frac{{\cal E}}{P}\right)\left(\frac{v_s}{c}\right)_S^2,
\label{gamma-1}
\end{equation}
where the subscript $S$ indicates the derivation at constant entropy and $v_s/c=\sqrt{(\partial  P/\partial  {\cal E})_S}$ is the speed of sound in units of speed of light. The  role of the adiabatic index in the stability condition is very important
and its effect will be  presented in detail  in the next section.

Finally, for convenience of the calculation recipe,  we reformulate   the instability condition~(\ref{cond-1}) in the compact form
\begin{equation}
\sigma^2T_{1L}=T_{1R}+T_{2R}+T_{3R}+T_{4R},
\label{Elab-1}
\end{equation}
where the correspondence  of the  terms $T_{iL(R)}$ with Eq.~(\ref{cond-1}) is obvious.

%

\section{Adiabatic index}
The adiabatic index is a basic ingredient of the instability criterion~(\ref{cond-1}). In particular, it is the quantity which incorporates all the basic characteristics of the equation of state on the instability formulae and consequently consists the {\it bridge } between the relativist structure of a spherical static object and the equation of state of the interior fluid. Specifically, for most of the equations of state of neutron star matter, $\gamma$ varies from 2 to 4. More precisely,  in some cases the adiabatic index is a weak function of the density while in other cases exhibits a more complicated density dependence~\cite{Haensel-07}.

The stability of a compact  object mainly depends on the interplay  between the equation of state of the fluid interior  and the strength of the relativistic field. In view of the above, the question which is naturally rises  is the following: Is it  possible, in the framework of stable configurations, to impose constraints on the equation of state of compact objects with respect to the values of the adiabatic index? And even more: Is it possible to  gain some useful information concerning extreme conditions (maximum mass, maximum central pressure and density and  e.t.c)  by studying the interplay between the adiabatic index (equation of state) and the stable configuration of a compact object?

The adiabatic index is  a function of the baryon density and consequently  exhibits  radial dependence on the instability criterion ~(\ref{cond-1}).
In a very few cases, as for example in a polytropic equation of state, the adiabatic index is a constant. Nevertheless, it is possible to define, via the instability criterion ~(\ref{cond-1}),  the averaged adiabatic index $\langle \gamma \rangle $ as~\cite{Merafina-89,Negi-2001}
\begin{equation}
\langle \gamma \rangle=\frac{\displaystyle \int_{0}^R e^{(\lambda+3\nu)/2}\frac{\gamma P}{r^2}\left(\frac{d}{dr}(r^2e^{-\nu/2}\xi)  \right)^2 dr  }
{ \displaystyle \int_{0}^R e^{(\lambda+3\nu)/2}\frac{P}{r^2}\left(\frac{d}{dr}(r^2e^{-\nu/2}\xi)  \right)^2 dr}.
\label{gamma-MV-1}
\end{equation}
In the special  case where $\xi=re^{\nu/2}$ the above formulae is considerably simplified and reduces to the expression
\begin{equation}
\langle \gamma \rangle=\frac{\displaystyle  \int_{0}^R e^{(\lambda+3\nu)/2}\gamma(r)P(r)r^2 dr}
{\displaystyle \int_{0}^R e^{(\lambda+3\nu)/2}P(r)r^2 dr}.
\label{gamma-MV-2}
\end{equation}
The averaged adiabatic index $\langle \gamma \rangle$ is a functional  both of the function $\xi$ as well as of the compactness parameter $\beta$.
Obviously,  $\langle \gamma \rangle$ is related to the condition of stability.  Chandrasekhar~\cite{Chandrasekhar-64} found that in the case of Schwarzschild constant-density interior solution, in the Newtonian limit,  in order to avoid the instability $\langle \gamma \rangle$ must satisfy  the inequality
\begin{equation}
\langle \gamma \rangle\geq \gamma_{cr}= \frac{4}{3}+\frac{19}{42}2\beta,
\label{gamma-Chns}
\end{equation}
where $\gamma_{cr}$ is the critical value of the adiabatic index (see discussion below).
Now, we will employ the approximation  that the adiabatic index $\gamma$ is a constant throughout the star~\cite{Chandrasekhar-64}. Obviously, this is  an artificial treatment since $\gamma$ (except in some specific cases) depends on the radius $r$. However, this approximation leads to a very useful insight and imposes  some  marginal constraints. In particular, this approximation directly  relates the equation of state, which characterizes the fluid, with a possible stable configuration. In view of the above consideration we will define the  critical value $\gamma_{cr}$ of the adiabatic index, as an effective  constant index which corresponds to neutral configuration $\sigma^2=0$. Now,  in the case where $\xi=re^{\nu/2}$,   $\gamma_{cr}$ is given by the equation
\begin{eqnarray}
\gamma_{cr}&=&\left[-4\int_{0}^R e^{(\lambda+3\nu)/2} r^3\frac{ dP}{dr} dr +\int_{0}^R e^{(\lambda+3\nu)/2}\left(\frac{d P}{d r}  \right)^2\frac{r^4}{P+ {\cal E}} dr \right.\nonumber \\
&-&\left.
\frac{8\pi G}{c^4}\int_0^R e^{3(\lambda+\nu)/2}P (P+{\cal E}) r^4 dr  \right]
\times\left(9\int_{0}^R e^{(\lambda+3\nu)/2} Pr^2 dr  \right)^{-1}.
\label{g-cr-1}
\end{eqnarray}
Obviously, the stability condition $\sigma^2 \geq 0$ leads to the inequality
\begin{equation}
\langle \gamma \rangle \geq \gamma_{cr}.
\label{ineq-gamma}
\end{equation}
The case $\langle \gamma \rangle = \gamma_{cr}$, as mentioned before,  corresponds to the onset of the instability. Finally, it should be noted that the adiabatic index defined in Eq.(\ref{gamma-1}), is a local characteristic of a specific equation of state and obviously is a function of the interior fluid density and consequently of the parameter of distance $r$. However, the effective adiabatic indices $\langle \gamma \rangle $ and $\gamma_{cr}$ have a global character
and they combine both characteristics of the equation of state as well as  characteristics of the relativistic field (see also Refs.~\cite{Bludman-73a,Bludman-73b,Ipser-70,Merafina-89,Herrera-89,Herrera-94,Yousaf-2015,Yousaf-2016}).

\section{Analytical solutions of Einstein's field equations}

\subsection{ Schwarzschild constant-density interior solution}
In the case of the  Schwarzschild interior solution, the density is constant through the star.
There is no physical justification of this solution, actually it is not a real one.  However, it is  very interesting since it combines the following:
a) in the interior of neutron stars the density is almost constant, b) it is very simple to allow an exact solution of Einstein's equations, c) it is useful for pedagogic reasons~\cite{Weinberg-72,Schutz-85}. In the present work we will use this solution in comparison with the three others. Actually,  the definition of the critical adiabatic index has been introduced by  Chandrasekhar by employing the constant interior solution~\cite{Chandrasekhar-64}. Briefly we present below the basic ingredients of the  Schwarzschild interior solution.
The metric functions are defined as
\begin{equation}
e^{-\lambda}=1-2\beta x^2, \qquad e^{\nu}=\left(\frac{3}{2}\sqrt{1-2\beta}-\frac{1}{2}\sqrt{1-2\beta x^2}    \right)^2, \quad x=r/R.
\label{metric-el-enu-uniform}
\end{equation}
The energy density and the pressure read
\begin{equation}
{\cal E}={\cal E}_c=\frac{3Mc^2}{4\pi R^3},
\label{Unif-E}
\end{equation}
\begin{equation}
\frac{P(x)}{{\cal E}_c}=\frac{\sqrt{1-2\beta}-\sqrt{1-2\beta x^2}}{\sqrt{1-2\beta x^2}-3\sqrt{1-2\beta }}.
\label{Unif-Pr}
\end{equation}
The ratio of the central pressure over central energy density $P_c/{\cal E}_c$, which plays important role on the stability condition is given by the expression
\begin{equation}
\frac{P_c}{{\cal E}_c}\equiv\frac{P(0)}{{\cal E}(0)}=\frac{\sqrt{1-2\beta}-1}{1-3\sqrt{1-2\beta }}.
\label{Unif-PcEc}
\end{equation}

According to  this solution, the central pressure becomes infinite when $\beta=4/9$ (actually, this upper limit holds for any
star~\cite{Buchdal-59}). The main drawback of this solution is the infinite value of the speed of sound.

\subsection{Tolman VII solution}
The Tolman VII solution is the most famous of the analytical ones. It is of great interest since it has the specific property the pressure and the density to vanish at the surface of the star. It has been extensively employed to neutron star studies while its  physical realization has been examined in detail very recently~\cite{Raghoonundun-15}. Actually, the stability of this solution has been examined by Negi {\it et al}.~\cite{Negi-1999,Negi-2001}. In the present work the stability is reexamined and in addition the results are compared with the other similar solutions.
The basic ingredients of the  Tolman VII solutions are presented below.
The metric functions are defined as follows
\begin{equation}
e^{-\lambda}=1-\beta x^2(5-3x^2), \qquad e^{\nu}=\left(1-\frac{5\beta}{3}\right)\cos^2\phi, \quad x=\frac{r}{R}
\label{metric-el-Tolm}
\end{equation}
where
\[\phi=\frac{w_1-w}{2}+\phi_1, \qquad \phi_1=\tan^{-1}\sqrt{\frac{\beta}{3(1-2\beta)}}\]
 and
 \[w=\ln\left(x^2-\frac{5}{6}+\sqrt{\frac{e^{-\lambda}}{3\beta}}\right), \quad w_1=\ln\left(\frac{1}{6}+\sqrt{\frac{1-2\beta}{3\beta}}  \right).  \]
The energy density and the pressure read
\begin{equation}
\frac{{\cal E}(x)}{{\cal E}_c}=(1-x^2), \quad {\cal E}_c=\frac{15Mc^2}{8\pi R^3},
\label{Tolm-E}
\end{equation}
\begin{equation}
\frac{P(x)}{{\cal E}_c}=\frac{2}{15}\sqrt{\frac{3e^{-\lambda}}{\beta }}\tan\phi-\frac{1}{3}+\frac{x^2}{5}.
\label{Tolm-Pr}
\end{equation}
There are some constraints related with the validity of the Tolman VII solution. In particular, the central value of pressure becomes infinite for $\beta=0.3862$, while the speed of sound remains less than the speed of light only for $\beta<0.2698$~\cite{Lattimer-2001}.

\subsection{Buchdahl solution}
We pay special effort to examine the stability of the Buchdahl solution. Firstly, Knutsen~\cite{Knutsen-88} found that this solution is unstable with respect to radial oscillations and consequently  must be discarded as a model for a star, where the temperature is essentially at absolute zero (this is the case of white dwarf and neutron star) or a convective equilibrium (that is a supermassive star). Later on, Negi~\cite{Negi-2007} showed that this solution is stable and also gravitationally bound for all  values of the compactness parameter. This analytical solution has been already used in the literature for studies related with the neutron star structure and may be a subject of  future similar studies. We consider that  it is worth to reexamine, with respect to the previous efforts,  the stability of this solutions.

The Buchdahl's solution has no particular physical basis. However,  it has have two specific properties: (i) it can be made casual everywhere in the star by demanding that the local speed of sound  is less than one  and   (ii) for small values of the pressure $P$ it reduces to ${\cal E}=12\sqrt{P^*P}$,  which, in the newtonian theory of stellar structure is the well known $n=1$ polytrope~\cite{Schutz-85}. So, Buchdahl's solution may be regarded as its relativistic generalization.
The equation of state, in Buchdahl's solution, has the following simple form
\begin{equation}
{\cal E}(P)=12\sqrt{P^{*}P}-5P,
\label{EOS-Buch}
\end{equation}
where $P$ is the local pressure and $P^{*}$ is a parameter.  In particular the metric functions are defined as
\begin{equation}
e^{\lambda(r')}=\frac{(1-2\beta)(1-\beta+u(r'))}{(1-\beta-u(r'))(1-\beta+\beta\cos(Ar'))^2},\qquad e^{\nu(r')}=\frac{(1-2\beta)(1-\beta-u(r'))}{1-\beta+u(r')},
\label{metric-el-Tolm}
\end{equation}
where
\[r'=\frac{r(1-2\beta)}{1-\beta+u(r')}, \qquad
u(r')=\beta\frac{\sin(Ar')}{Ar'},\qquad  A^2=\frac{288\pi P^{*}G}{c^4 (1-2\beta)}.  \]
The energy density and the pressure read
\begin{equation}
\frac{{\cal E}(r')}{{\cal E}_c}=\frac{(2-2\beta-3u(r'))}{(2-5\beta)(1-\beta+u(r'))^2}\frac{u(r')}{\beta},
\label{Ec-Buch}
\end{equation}
\begin{equation}
\frac{P(r')}{{\cal E}_c}=\frac{\beta}{(1-\beta+u(r'))^2(2-5\beta)} \left(\frac{u(r')}{\beta}\right)^2,
\label{Pr-Buch}
\end{equation}
where
\begin{equation}
P_c=36P^*\beta^2, \qquad {\cal E}_c=72 P^*\beta(1-5\beta/2), \qquad \frac{P_c}{{\cal E}_c}=\frac{\beta}{2-5\beta}.
\label{ins-buc-3}
\end{equation}
It is more convenient to use the variable $x'=r'/R$  instead of   $x=r/R$ and in this case we have
\begin{equation}
x=\frac{1-\beta+u(x')}{1-2\beta}x',
\label{ins-buc-2-a}
\end{equation}
where
\begin{equation}
u(x')=\beta\frac{\sin(AR x')}{AR x'}, \qquad AR=\frac{1-\beta}{1-2\beta}\pi.
\label{ins-buc-2}
\end{equation}
The Buchdahl solution is physically meaningful  in restricted domains~\cite{Lattimer-2001}. These are:  a) in order to ensure that ${\cal E}(r)>0$ then must $\beta < 0.4$, b) the causality condition $v_s<c$ demands that $\beta<1/6$ and c) the condition $v_s^2>0$ satisfied only when $\beta<1/5$.

The upper limit of $x'$ which corresponds to the surface of the fluid sphere is given by \[ x'_{max}=\frac{1-2\beta}{1-\beta}\] and the various terms of the Chandrasekhar instability criterion, considering that $\tilde{\xi}=\xi/r$,  take the form
\begin{eqnarray}
T_{1L}&=& R^5{\cal E}_c\int_{0}^{x'_{max}} e^{(3\lambda-\nu)/2}\left(\frac{P(x')}{{\cal E}_c}+\frac{{\cal E }(x')}{{\cal E}_c}\right) \left(\frac{1-\beta+u(x')}{1-2\beta}x'\right)^2 \tilde{\xi}^2 \nonumber \\
&\times& \frac{1}{1-2\beta}\left[(1-\beta)+x'\frac{du(x')}{dx'}+u(x')\right]dx',
\label{T1L-3-Buh}
\end{eqnarray}
\begin{equation}
T_{1R}=4R^3{\cal E}_c\int_{0}^{x'_{max}} e^{(\lambda+\nu)/2} \left(\frac{1-\beta+u(x')}{1-2\beta}x'\right) \frac{ d(P(x')/{\cal E}_c)}{dx'}\tilde{\xi}^2 dx',
\label{T1R-3-Buh}
\end{equation}
\begin{eqnarray}
T_{2R}&=&R^3{\cal E}_c \int_{0}^{x'_{max}} e^{(\lambda+3\nu)/2}\gamma(x')  \frac{P(x')}{{\cal E}_c}  \left(\frac{1-\beta+u(x')}{1-2\beta}x'  \right)^{-2} \nonumber
\\&\times& \left(\left(\frac{1}{1-2\beta}\left[(1-\beta)+x'\frac{du(x')}{dx'}+u(x')\right]\right)^{-1}\frac{d}{dx'}
\left[  \left(\frac{1-\beta+u(x')}{1-2\beta}x'\right)^2 e^{-\nu/2}\tilde{\xi}  \right]\right)^2 \nonumber \\
&\times&\frac{1}{1-2\beta}\left[(1-\beta)+x'\frac{du(x')}{dx'}+u(x')\right]dx',
\label{T2R-2-Buh}
\end{eqnarray}
\begin{eqnarray}
T_{3R}&=&-R^3{\cal E}_c\int_{0}^{x'_{max}} e^{(\lambda+\nu)/2}\left(\left(\frac{1}{1-2\beta}\left[(1-\beta)+x'\frac{du(x')}{dx'}+u(x')\right]  \right)^{-1}          \frac{d (P(x')/{\cal E}_c)}{d x'}  \right)^2 \nonumber \\
&\times& \left(\frac{1-\beta+u(x')}{1-2\beta}x'  \right)^2    \frac{\tilde{\xi}^2}{\frac{P(x')}{{\cal E}_c}+\frac{{\cal E}(x')}{{\cal E}_c}} \frac{1}{1-2\beta}\left[(1-\beta)+x'\frac{du(x')}{dx'}+u(x')\right]dx',
\label{T3R-3-Buh}
\end{eqnarray}
\begin{eqnarray}
T_{4R}&=&\frac{2\pi^2\beta(1-\beta)^2(1-5\beta/2)}{1-2\beta}{\cal E}_cR^3\int_0^{x'_{max}} e^{(3\lambda+\nu)/2}\frac{P(x')}{{\cal E}_c} \left(\frac{P(x')}{{\cal E}_c}+\frac{{\cal E}(x')}{{\cal E}_c}\right) \nonumber \\
&\times&  \left(\frac{1-\beta+u(x')}{1-2\beta}x'  \right)^2\tilde{\xi}^2 \frac{1}{1-2\beta}\left[(1-\beta)+x'\frac{du(x')}{dx'}+u(x')\right]dx'.
\label{T4R-3-Buch}
\end{eqnarray}


\subsection{ Nariai IV solution}
The Nariai IV solution~\cite{Nariai-50,Nariai-51,Nariai-99}  is the most complicated and less known, compared to the previous ones.
This solution has been employed in  neutron stars  studies~\cite{Lattimer-05a,Moustakidis-016}.
However,  the stability of this solution, according to our knowledge, is examined for a first time in the literature.
In the present work we will employ the parametrization used in Ref.~\cite{Lattimer-05}, since it is more convenient  and with  obvious physical representation. Due to the complicated character of this solution, the  presentation is given below with all the details. In particular, the  metric functions are
defined as
\begin{equation}
e^{-\lambda(r')}=\left(1-\sqrt{3\beta}\left(\frac{r'}{R'} \right)^2 \tan f(r') \right)^2, \qquad \qquad e^{\nu(r')}=(1-2\beta)\frac{E^2}{C^2}\left( \frac{\cos g(r')}{\cos f(r')} \right)^2,
\label{el-en-Nariai}
\end{equation}
where
\begin{equation}
r=\frac{E}{C}\frac{r'}{\cos f(r')}\sqrt{1-2\beta}, \quad R'=\frac{R C}{\sqrt{1-2\beta}},
\label{Nariai-4}
\end{equation}
and
\begin{equation}
f(r')=\cos^{-1}E+\sqrt{\frac{3\beta}{4}}\left[1-\left(\frac{r'}{R'} \right)^2\right], \quad
g(r')=\cos^{-1}C+\sqrt{\frac{3\beta}{2}}\left[1-\left(\frac{r'}{R'} \right)^2\right],
\label{Nariai-1-1}
\end{equation}
\begin{equation}
E^2=\cos^2f(R')=\frac{2+\beta+2\sqrt{1-2\beta}}{4+\beta/3}, \quad
C^2=\cos^2 g(R')=\frac{2E^2}{2E^2+(1-E^2)(7E^2-3)^2(5E^2-3)^{-2}}.
\label{Nariai-2}
\end{equation}
The energy density  ${\cal E}(r')$ and the pressure $P(r')$ are  expressed in terms of  the parametric variable $r'$
\begin{eqnarray}
{\cal E}(r')=\frac{\sqrt{3\beta}}{4\pi R'^2(1-2\beta)}\frac{C^2}{E^2}\frac{c^4}{G}\left[3\sin f(r')\cos f(r')-\sqrt{\frac{3\beta}{4}}\left(\frac{r'}{R'}\right)^2(3-\cos^2f(r'))
\right],
\label{Nariai-1-den}
\end{eqnarray}
\begin{eqnarray}
P(r')&=&\frac{\sqrt{3\beta}\cos f(r')}{4\pi R'^2(1-2\beta)}\frac{C^2}{E^2}\frac{c^4}{G}\left[ \sqrt{2}\cos f(r')\tan g(r')\left(1-\sqrt{3\beta}\left(\frac{r'}{R'} \right)^2\tan f(r')  \right) \right. \nonumber \\
&-&\left.\sin f(r')\left(2-\frac{3}{2}\sqrt{3\beta}\left(\frac{r'}{R'} \right)^2\tan f(r')   \right)   \right].
\label{Nariai-1-pres}
\end{eqnarray}
In addition the central energy density and pressure are
\begin{eqnarray}
{\cal E}(0)\equiv {\cal E}_c=\frac{\sqrt{3\beta}}{4\pi R'^2(1-2\beta)}\frac{C^2}{E^2} 3\frac{c^4}{G}\sin f(0)\cos f(0),
\label{Nariai-1-denden-c}
\end{eqnarray}
\begin{eqnarray}
P(0)\equiv P_c=\frac{\sqrt{3\beta}}{4\pi R'^2(1-2\beta)}\frac{C^2}{E^2} \frac{c^4}{G}\sin f(0)\cos f(0)\left(\sqrt{2}\cot f(0)\tan g(0)-2   \right),
\label{Nariai-1-pres-c}
\end{eqnarray}
and also
\begin{equation}
\frac{P_c}{{\cal E}_c}=\frac{1}{3}\left(\sqrt{2}\cot f(0)\tan g(0)-2   \right).
\label{ratio-Pc/ec}
\end{equation}
The central pressure remains finite only when $\beta<0.4126$, while the causality is satisfied when $\beta<0.2277$~\cite{Lattimer-05}.
It is more convenient now to use the variable $x'=r'/R$  instead  of the variable $x=r/R$.

The upper limit of $x'$, which corresponds to the surface of the fluid sphere, is given by \[ x'_{max}=\frac{R'}{R}=\frac{C}{\sqrt{1-2\beta}}\] and the various terms of the Chandrasekhar instability criterion are reformulated as follows
\begin{eqnarray}
T_{1L}&=& R^5{\cal E}_c\int_{0}^{x'_{max}} e^{(3\lambda-\nu)/2}\left(\frac{P(x')}{{\cal E}_c}+\frac{{\cal E }(x')}{{\cal E}_c}\right) \left(\frac{E}{C}\frac{x'}{\cos f(x')}\sqrt{1-2\beta}\right)^2 \tilde{\xi}^2 \nonumber \\
&\times& \frac{E\sqrt{1-2\beta}}{C}\left[\frac{1}{\cos f(x')}-\sqrt{3\beta}\left(\frac{x'}{x'_{max}} \right)^2\frac{\sin f(x')}{\cos^2 f(x')}\right]dx',
\label{T1L-3-Nariai}
\end{eqnarray}
\begin{equation}
T_{1R}=4R^3{\cal E}_c\int_{0}^{x'_{max}} e^{(\lambda+\nu)/2}\left(\frac{E}{C}\frac{x'}{\cos f(x')}\sqrt{1-2\beta}  \right)  \frac{ d(P(x')/{\cal E}_c)}{dx'}\tilde{\xi}^2 dx',
\label{T1R-3-Nariai}
\end{equation}
\\
\begin{eqnarray}
T_{2R}&=&R^3{\cal E}_c \int_{0}^{x'_{max}} e^{(\lambda+3\nu)/2}\gamma(x')  \frac{P(x')}{{\cal E}_c}  \left(\frac{E}{C}\frac{x'}{\cos f(x')}\sqrt{1-2\beta}\right)^{-2} \nonumber
\\&\times& \left(\left(\frac{E\sqrt{1-2\beta}}{C}\left[\frac{1}{\cos f(x')}-\sqrt{3\beta}\left(\frac{x'}{x'_{max}} \right)^2\frac{\sin f(x')}{\cos^2 f(x')}\right]       \right)^{-1}\frac{d}{dx'}
\left[  \left(\frac{E}{C}\frac{x'}{\cos f(x')}\sqrt{1-2\beta}      \right)^2 e^{-\nu/2}\tilde{\xi}  \right]\right)^2 \nonumber \\
&\times&\frac{E\sqrt{1-2\beta}}{C}\left[\frac{1}{\cos f(x')}-\sqrt{3\beta}\left(\frac{x'}{x'_{max}} \right)^2\frac{\sin f(x')}{\cos^2 f(x')}\right] dx',
\label{T2R-2-Buh}
\end{eqnarray}
\\
\begin{eqnarray}
T_{3R}&=&-R^3{\cal E}_c\int_{0}^{x'_{max}} e^{(\lambda+\nu)/2}\left(\left(\frac{E\sqrt{1-2\beta}}{C}\left[\frac{1}{\cos f(x')}-\sqrt{3\beta}\left(\frac{x'}{x'_{max}} \right)^2\frac{\sin f(x')}{\cos^2 f(x')}\right]      \right)^{-1}          \frac{d (P(x')/{\cal E}_c)}{d x'}  \right)^2 \nonumber \\
&\times& \left(\frac{E}{C}\frac{x'}{\cos f(x')}\sqrt{1-2\beta} \right)^2    \frac{\tilde{\xi}^2}{\frac{P(x')}{{\cal E}_c}+\frac{{\cal E}(x')}{{\cal E}_c}}
\frac{E\sqrt{1-2\beta}}{C}\left[\frac{1}{\cos f(x')}-\sqrt{3\beta}\left(\frac{x'}{x'_{max}} \right)^2\frac{\sin f(x')}{\cos^2 f(x')}\right]
dx',
\label{T3R-3-Nariai}
\end{eqnarray}
\\
\begin{eqnarray}
T_{4R}&=&\frac{\sqrt{3\beta}}{E^2}\sin f(0)\cos f(0)     {\cal E}_cR^3\int_0^{x'_{max}} e^{(3\lambda+\nu)/2}\frac{P(x')}{{\cal E}_c} \left(\frac{P(x')}{{\cal E}_c}+\frac{{\cal E}(x')}{{\cal E}_c}\right) \nonumber \\
&\times&  \left(\frac{E}{C}\frac{x'}{\cos f(x')}\sqrt{1-2\beta}  \right)^2\tilde{\xi}^2 \frac{E\sqrt{1-2\beta}}{C}\left[\frac{1}{\cos f(x')}-\sqrt{3\beta}\left(\frac{x'}{x'_{max}} \right)^2\frac{\sin f(x')}{\cos^2 f(x')}\right]dx'.
\label{T4R-3-Nariai}
\end{eqnarray}

\section{Results and discussion}
A basic property characterizes the three solutions,  the derived pressure and energy density of the fluid vanish at the surface. All these properties are displayed in Fig.~1a and b. The plots correspond to the case of  $M=1.4 M_{\odot}$, $R=12.5 \ {\rm Km}$ and $\beta=0.1652$. This mass-radius pair is consistent  with the observations data for neutron stars while the corresponding value of the compactness satisfies all the relevant constraints.

In Fig.~2a we display the dependence of the speed of sound (at the center of the star)  on the compactness, for the three solutions. The violation of the causality is also indicated in each case. The dependence of the ratio $P_c/{\cal E}_c$ on the compactness is presented in Fig.~2b. Obviously, this dependence is similar for the three  cases and up to high values of $\beta$. This behavior will be discussed in  detail  bellow.

In Fig.~3, we present a mass-radius  diagram, for the Tolman VII solution,  where  the constraints imposed  on the compactness parameter $\beta$ in order  a) to satisfy   the causality, b) to ensure  of stability and c) to ensure finite value of the central pressure (and/or speed of sound) have been indicated. Obviously, the constraints on $\beta$ impose also constraints on the maximum mass (for a fixed value of the radius) and on the minimum radius (for a fixed mass). The above constraints, for each solution, have been  summarized also in Table~1.

In Fig.~4a we display the dependence of the frequency of the pulsations $(\sigma R)^2$ on the compactness parameter $\beta=GM/Rc^2$  for the three analytical solutions. The approximation (\ref{aprox-sigma}) is also included (see below). We employ, in each case, the {\it trial} function $\xi=re^{\nu/4}$. The Tolman VII solution becomes unstable for $\beta=0.3428$ (confirming  the value that found also by~\cite{Negi-1999}) and still remains unstable up to  the value $\beta=0.3862$ where the ratio $P_c/{\cal E}_c$ remains finite. The Nariai IV solution is stable up to the value $\beta=0.4126$ where the ratio  $P_c/{\cal E}_c$ becomes infinite. The Buchdahl solution is stable up to the value $\beta=0.2$ since for $\beta>0.2$ the speed of sound becomes $v_s^2<0$.   It is worth to noticing  that all the analyzed solutions are stable even in the region of $\beta$ where the causality condition is violated (see Section~5 for more details). Nevertheless, the mentioned violation does not lead to instability of the compact object (see also Ref.~{\cite{Mak-2013}).  However, in order to compare the results with those originated from realistic equation of states the causality constraints must be taken into account (see also the discussion below).

We concentrate now on the stability of the Buchdahl solution. We performed  a Taylor expansion of the instability  condition (\ref{cond-1}) and we founded, for low values of $\beta$, the approximation
\begin{equation}
(\sigma R)^2\simeq \frac{3\pi^2}{2(\pi^2-6)}\beta.
\label{aprox-sigma}
\end{equation}
The above approximation confirms the stability of the Buchdahl solution and also reproduces very well the numerical results up to the value $\beta=0.1$.
Moreover,  in Fig.~4b  we present the $(\sigma R)^2$ dependence on $\beta$  for three different {\it trial} functions $\xi$. The analytical approximation (\ref{aprox-sigma}) has been also included for comparison. The dependence $(\sigma R)^2$ on $\beta$ is almost insensitive on the choice of  the trial function $\xi $, especially for  low values of $\beta$. The present results, are very close to the results of Negi~\cite{Negi-2007} and we confirm his statement that the Buchdahl solution is stable for all values of $\beta$. In order to provide more details of our study, we present  in Table~2 the numerical results concerning the Buchdahl solution in comparison with the results of Negi~\cite{Negi-2007} and for future studies.

In Fig.~5a we plot the averaged adiabatic index $\langle \gamma \rangle$ and the corresponding critical adiabatic index $\gamma_{cr}$ for the three analytical solutions as a function of the compactness $\beta$. In addition, the result of the uniform density for  $\gamma_{cr}$, and also   the approximation $\gamma_{cr}=4/3+19/21\beta$ (introduced by Chandresekhar~\cite{Chandrasekhar-64}) are also included. In the Newtonian limit  $\beta \rightarrow 0 $, all the cases reproduce the expected result that is  $\gamma_{cr}\simeq 4/3$, while in the post-Newtonian approximation dynamical stability requires that $\langle \gamma \rangle\ > \gamma_{cr}$.

Fig.~5a demonstrates the stability of the analytical solutions since in each case  the condition $\langle \gamma \rangle\ >\gamma_{cr}$ is ensured for the relevant range  of $\beta$. The only exception is the Tolman VII solution where for $\beta>0.3428$, the inequality $\langle \gamma \rangle\ <\gamma_{cr}$ takes place,   ensuring the instability  of the solution in the mentioned region (see also Fig.~4a).
It should be noted  that  the dependence of $\langle \gamma \rangle$ on $\beta$ is model dependent. However, the  most distinctive feature indicated in Fig.~5a is a similar  dependence of $\gamma_{cr}$ on $\beta$, for all the analytical models,  even for high values of $\beta$. The above result leads to the   conclusion that $\gamma_{cr}$ exhibits a similar behavior, at least for the configurations originated from analytical solutions of the relativistic equations. However, the above statement must  be  considered for the case where realistic equations of state are used for the construction of spherical relativistic configurations.
Using the Tolman VII solution as a guide of the mentioned relations we found that the expression
\begin{equation}
\gamma_{cr}=C_0+C_1 e^{\beta/C_2}, \qquad C_0=1.2575, \ C_1=0.0873, \ C_2=0.1119
\label{gamm-fit}
\end{equation}
reproduces very well the numerical results of  the Tolman VII solution,  up to the value $\beta=0.2698$ (which corresponds to the causality limit) and also the results of the other analytical solutions.  It is also natural  to extent the above conjecture and to consider that the dependence~(\ref{gamm-fit}) can be generalized to include also the adiabatic index which corresponds to realistic equation of state. In addition, we consider also the possibility,  to impose some constraints on the value of the adiabatic index in order to ensure the stability of a very compact object (dwarf with mass close to the limiting mass, neutron stars and  supermassive stars too).

In order to clarify further the effects of the equation of state, via the adiabatic index, on the stability condition,  we plot in Fig.~5b the dependence of the averaged adiabatic index $\langle \gamma \rangle$ and  the critical adiabatic index $\gamma_{cr}$ on the ratio $P_c/{\cal E}_c$ for the four analytical solutions. The Chandrasekhar
approximation~\cite{Chandrasekhar-64}
\begin{equation}
\gamma_{cr}=\frac{4}{3}+\frac{19}{42}\left(1-\left(\frac{1+P_c/{\cal E}_c}{1+3P_c/{\cal E}_c} \right)^2 \right),
\label{gcr-pcec-uni}
\end{equation}}
have been also included for comparison. The most distinctive feature is the occurrence of an almost linear dependence between $\gamma_{cr}$ and $P_c/{\cal E}_c$ in all the cases.
In particular, this dependence is of the form
\begin{equation}
\gamma_{cr}=\frac{4}{3}+{\cal K}\left(\frac{P_c}{{\cal E}_c}\right),
\label{g-PcEc-1}
\end{equation}
where the values of the parameter  ${\cal K}$ for the Tolman VII, Buchdahl and  Nariai IV  are $2.2852$, $2.1082$ and  $2.5904$ respectively (see also Table~1).
Actually, the values of $\gamma_{cr}$ provide, for each solution, the lower limit which correspond to stable configurations.
We note that  in all  cases the upper bound $P_c/{\cal E}_c$ is taken in order to ensure the causality of the central value of the speed of sound. Moreover, for visual clarity,    the dependence of the critical and averaged adiabatic indices  on $\beta$ and $P_c/{\cal E}_c$  have been displayed  separately in Fig.~6.   

In the uniform density solution, since there is no restriction  from the speed of sound,  we can extend the study even for higher values of $\beta$ and consequently for the ratio $P_c/{\cal E}_c$. Considering, for physical reasons, as an upper bound the value of $P_c/{\cal E}_c=1$, we found that expression (\ref{g-PcEc-1}) holds in a very good accuracy. However, the values of ${\cal K}$ exhibit moderate dependence  on the form of the trail function $\xi$, for high values of $P_c/{\cal E}_c$. In particular, we found that ${\cal K}=1.8734$  for  $\xi(r)=r e^{\nu/4}$  and  ${\cal K}=1.998$ for the $\xi(r)=r e^{\nu/2}$.
We conjecture that the Schwarzschild  interior solution marks the absolute lower limit to the critical adiabatic index, given by the expression
\begin{equation}
\gamma_{cr}^{min}=\frac{4}{3}+1.8734\left(\frac{P_c}{{\cal E}_c}\right).
\label{g-PcEc-abs}
\end{equation}
Consequently, for any analytical or numerical solution of the TOV equation and for a given value of $P_c/{\cal E}_c$, it always holds
\begin{equation}
\langle \gamma \rangle > \gamma_{cr}^{min}.
\label{abs-ineq}
\end{equation}
The universality of the  expression (\ref{g-PcEc-1}), mainly for low values of $P_c/{\cal E}_c$,   is essential in order to relate the present results with those originating from realistic equations of state. Moreover, it is of significant interest to examine, in the framework of realistic equations of state, in which extent the condition of neutral configuration $\langle \gamma \rangle=\gamma_{cr}$ satisfies the expression (\ref{g-PcEc-1}).

In any case, useful insight can be gained from the comparison of the results originating from  numerical and analytical solution of the TOV equations. For example,   the stable configurations, which correspond to a numerical solution of the TOV equation using realistic equation of state  will satisfy the inequality $\langle \gamma \rangle>\gamma_{cr}$. Now, from  Eqs.(\ref{gamma-MV-2}) and (\ref{g-PcEc-1}) we get the  deduced  stability inequality
\begin{equation}
\int_{0}^R e^{(\lambda+3\nu)/2}P(r)r^2\left(\gamma(r)-\frac{4}{3}-{\cal K}\left(\frac{P_c}{{\cal E}_c}\right)  \right)dr>0.
\label{cond-eos-1}
\end{equation}
In general, the adiabatic index $\gamma$, which corresponds to any realistic equation of state,  is a function of the density and consequently of the distance. However, in some cases $\gamma $ is a weak function of the density~\cite{Haensel-07,Douchin-01}. In these cases,  considering that the adiabatic index is almost a  constant,   the inequality (\ref{cond-eos-1}) leads to the following simple constrain, in order to ensure the stability of the configuration
\begin{equation}
\frac{P_c}{{\cal E}_c} \leq \frac{1}{ {\cal K}}\left(\gamma-\frac{4}{3}\right).
\label{Eos-cons-2}
\end{equation}
The inequality (\ref{Eos-cons-2}) holds in a very good accuracy mainly for $P_c/{\cal E}_c<0.25$ where the value of ${{\cal K}}$ is quite well fixed. For higher values of $P_c/{\cal E}_c$ one has to impose the proper  uncertainties on the values of ${{\cal K}}$. In any case, it is interesting that the onset of instability exhibits so simple dependence on the ratio $P_c/{\cal E}_c$ even for the case of relativistic stars. In Fig.~7 we indicate the stability (instability) window which is defined by the analytical solutions employed in the present study. The upper curve is the absolute upper limit given by the expression
\begin{equation}
\left(\frac{P_c}{{\cal E}_c}\right)_{{\rm max}}=0.5338\left(\gamma-\frac{4}{3}\right)
\label{Eos-cons-3}
\end{equation}
which corresponds to Eq.~(\ref{Eos-cons-2}) with  ${\cal K}=1.8734$ (originated from the  Schwarzschild constant-density interior solution). The  lower curve corresponds to the value  ${\cal K}=2.5904$ which originated from the Nariai IV solution. The relative uncertainty indicated  also by the intermediate shaded region.

Finally, it is worth to pointing out that  there is a simple and practical  criterion which can be obtained assuming that the adiabatic index is the same as in a slowly deformed matter~\cite{Haensel-07}. In this case the model is stable when the inequality $dM/d{\cal E}_c>0$ holds. However, the above condition is just necessary but not sufficient, as stated for the criterion  $(\ref{cond-1})$.

\section{Concluding remarks}
The stability of three analytical solutions, suitable to describe compact relativistic objects,  has been examined. The stability in all the cases is ensured for a large range of the compactness parameter $\beta$.
Our study, concerning the stability of the Buchdahl solution,  supports the finding of Negi~\cite{Negi-2007} that the solution is stable  and is in contradiction with the corresponding statement of Knutsen~\cite{Knutsen-88}.
Our study leads to the conclusion that the   Nariai IV solution is also stable and suitable for physical applications. According to our knowledge the stability of this solution has never been examined in the past. We found that the critical adiabatic index exhibits a model independent behavior  on $\beta $ even for high values of the compactness and also an almost linear and  model independent behavior  on the ratio  $P_c/{\cal E}_c$.
This relation holds  even for higher values of the compactness ($\beta\simeq 0.2$) and the ratio $P_c/{\cal E}_c$ ($P_c/{\cal E}_c \simeq 0.25$).
A planned future work will extend the study in order to include results originating from the use of realistic equations of state.
In this case, it will be possible to impose additional constraints on realistic equations of state of the fluid interior, by demanding stable  configurations for relativistic stars.

\section*{Acknowledgments}
This work was supported  by the Aristotle University of Thessaloniki Research Committee under Contract No. 89286.

 \begin{table}[h]
\begin{center}
\caption{ The values of the parameter $\beta$ (and the corresponding ratios $P_c/{\cal E}_c$) which imposed in order a) to satisfy the  causality, b)  to ensure  the stability and c) to ensure finite value of the central pressure (and/or speed of sound) for each analytical solution. The values of the fitting parameter {\cal K} (employed in Eq.~(55)),  have been included too (for more details see text).  }
 \label{t:1}
\vspace{0.5cm}
\begin{tabular}{|c|c|c|c|c|c|}
\hline
   &  causality     & finite ratio $P_c/{\cal E}_c$ & stability  & finite  speed of sound   &       \\
\hline
Solution     &  $\beta$ ($P_c/{\cal E}_c$) & $\beta$ ($P_c/{\cal E}_c$) & $\beta$ ($P_c/{\cal E}_c$)  & $\beta$ ($P_c/{\cal E}_c$) & {\cal K}           \\
\hline
\hline
Tolman VII           &   0.2698 (0.382) &  0.3862 $(\infty)$  &  0.3428 (1.19)    & 0.3862 $(\infty)$  &     2.2852                \\

Buchdahl           &     0.1667 (0.143)   &  0.4000 $(\infty)$   &   -      & 0.2000 (0.2)  &    2.1082               \\

Nariai IV            &   0.2277 (0.245)  &   0.4126 $(\infty)$   &   -     &  0.4126$(\infty)$   &      2.5904        \\

 \hline
 \end{tabular}
\end{center}
\end{table}

 \begin{table}[h]
\begin{center}
\caption{ The quantity $(\sigma R)^2_i, \ i=1,2,3$, for the Buchdahl  solution,  as a function of the compactness parameter $\beta$ for three cases. The case $1$ corresponds to the trial function  $\xi(r)=r e^{\nu/4}$, the case $2$ to  $\xi(r)=r e^{\nu/2}$ and case $3$ to $\xi(r)=b_1r\left(1+a_1r^2+2_2r^4+a_3r^6\right)e^{\nu/2}$ . }
 \label{t:1}
\vspace{0.5cm}
\begin{tabular}{|c|c|c|c|c|c|}
\hline
 { $\beta$}     &  $(\sigma R)_1^2$   & $(\sigma R)_2^2$  &$(\sigma R)_3^2$             \\
\hline
\hline
0.00005            & 0.00019    & 0.00019     &   0.00019                     \\

0.00050            & 0.00191    &  0.00191       &  0.00189                      \\

0.00500            & 0.01909     &  0.01908   &    0.01881                  \\

0.01000            & 0.03809    & 0.03807     &   0.03748                     \\

0.02000            & 0.07595    &  0.07586       &  0.07450                     \\

0.03000            & 0.11372     &  0.11351   &    0.11117                   \\

0.04000            & 0.15160    & 0.15119     &   0.14763                    \\

0.05000            & 0.18980    &  0.18911       &  0.18401                     \\

0.06000            & 0.22853    &  0.22750       &  0.22052                     \\

0.07000            & 0.26818    &  0.26664       &  0.25736                    \\

0.08000            & 0.30911     &  0.30691   &    0.29482                  \\

0.09000            & 0.35183    & 0.34877     &   0.33323                    \\

0.10000            & 0.39698    &  0.39282       &  0.37303                      \\

0.11000            & 0.44542     &  0.43981   &    0.41480                 \\

0.12000            & 0.49833     &  0.49082   &    0.45928                  \\

0.13000            & 0.55733    &  0.54731     &   0.50754                   \\

0.14000            & 0.62485    &  0.61145       &  0.56113                    \\

0.15000            & 0.70462    &  0.68655       &  0.62236                     \\

0.16000            & 0.80281    &  0.77809       &  0.69506                     \\

0.17000            & 0.93080     &  0.89607   &    0.78609                  \\

0.18000            & 1.11299    & 1.06182     &   0.90994                    \\

0.19000            & 1.42006    &  1.33655       &  1.10755                     \\

0.19995            & 2.63092     &  2.38451   &    1.81363               \\

 \hline
 \end{tabular}
\end{center}
\end{table}


\vspace{3cm}
\begin{figure}
\centering
\includegraphics[height=8.5cm,width=8.5cm]{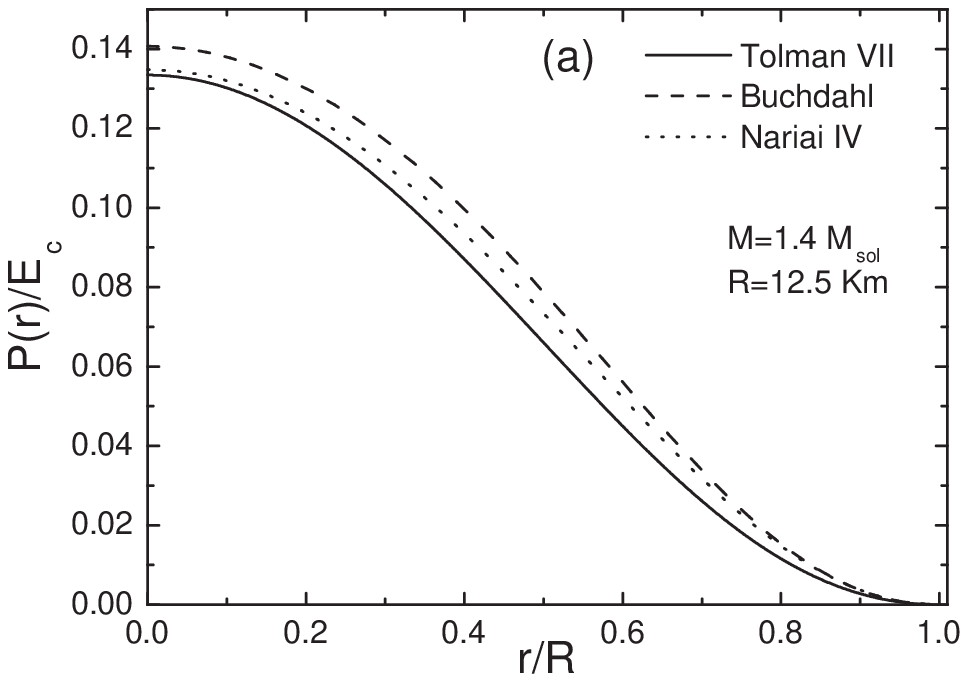}\
\includegraphics[height=8.5cm,width=8.5cm]{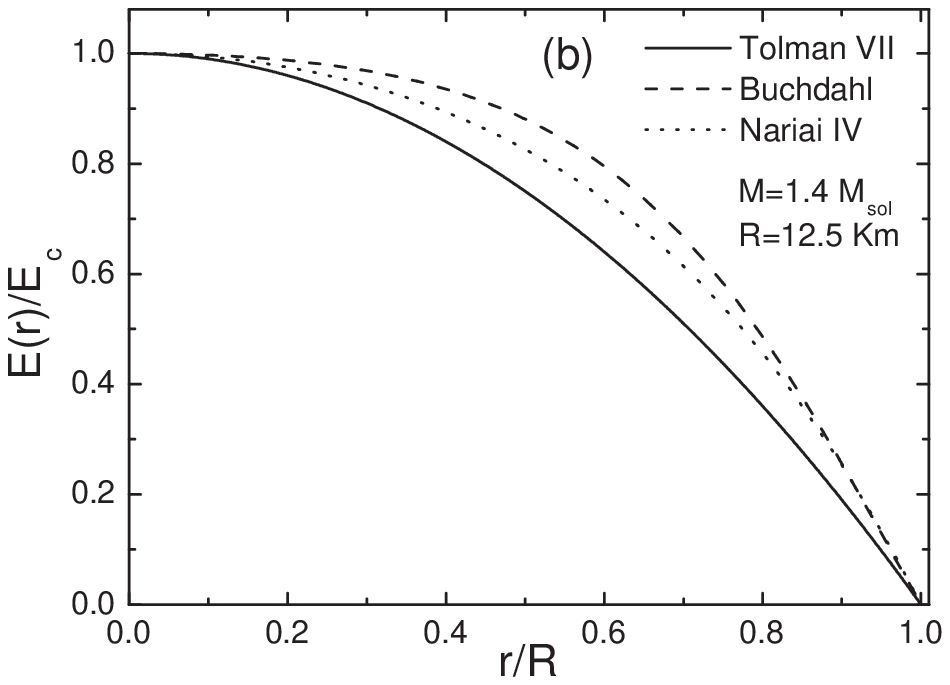}\
\caption{ (a) The pressure (in units of  ${\cal E}_c$)  and (b) the energy density ${\cal E}$ (in units of  ${\cal E}_c$) as a functions of the radius for the three analytical
solutions, Tolman VII, Buchdahl and Nariai IV and for the specific case of $M=1.4 M_{\odot}$ and $R=12.5 \ {\rm Km}$.   } \label{Esnm-pnm}
\end{figure}
\newpage
\begin{figure}
 \centering
\includegraphics[height=8.5cm,width=8.5cm]{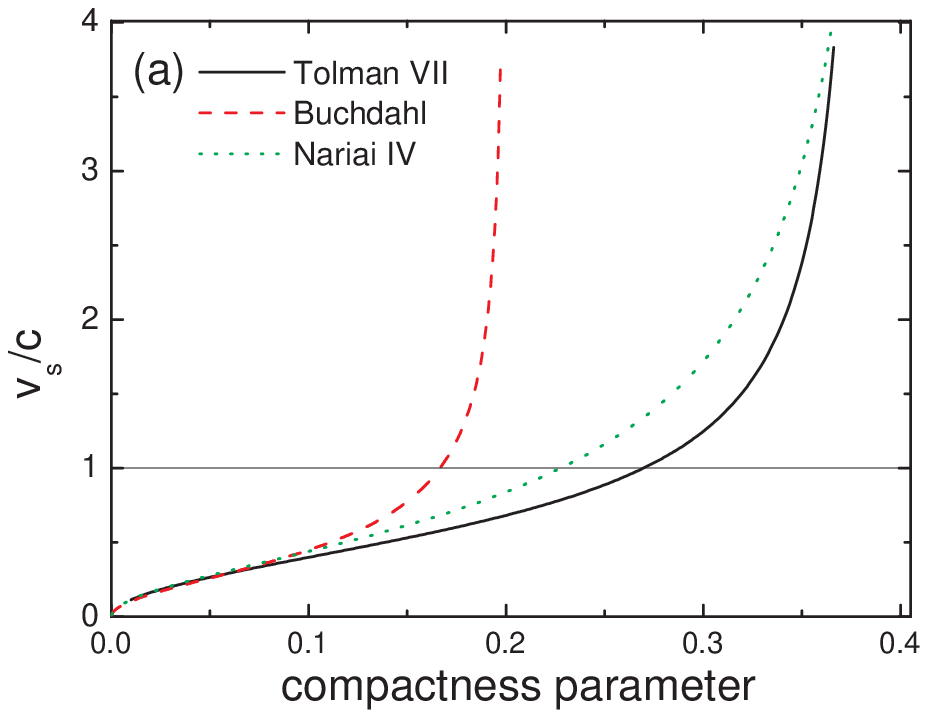}\
\includegraphics[height=8.5cm,width=8.5cm]{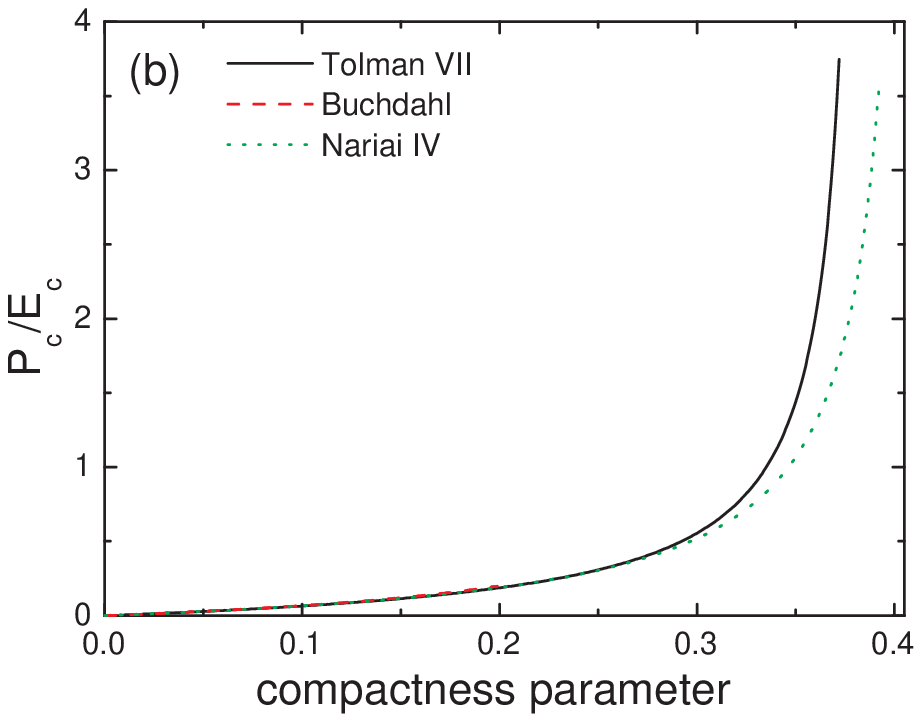}\
\caption{ (a) The  value of the speed of sound (at the center of the star) and (b) the ratio $P_c/{\cal E}_c$ as  a functions of the
compactness parameter $\beta$  of  the three analytical solutions.   } \label{Esnm-pnm}
\end{figure}

\begin{figure}
 \centering
\includegraphics[height=8.1cm,width=9.5cm]{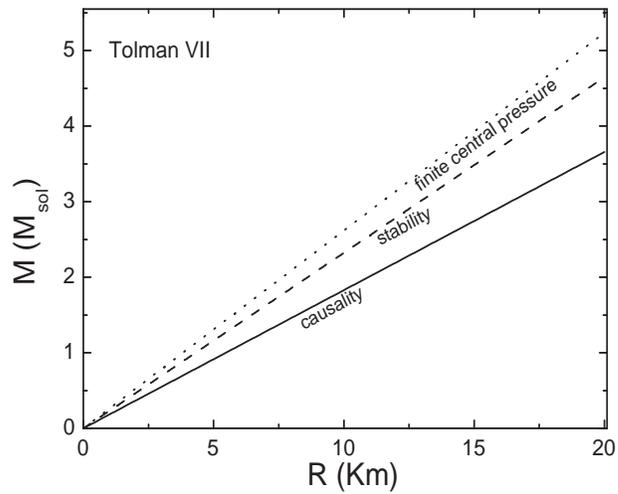}\
\caption{The mass-radius  diagram for the Tolman VII solution. The  constraints imposed  on the compactness parameter $\beta$ are applied in order  a) to satisfy    the causality, b) to ensure  of stability and c) to ensure finite value of the central pressure (and/or speed of sound) have been indicated.  }
\end{figure}

\begin{figure}
 \centering
\includegraphics[height=8.5cm,width=8.5cm]{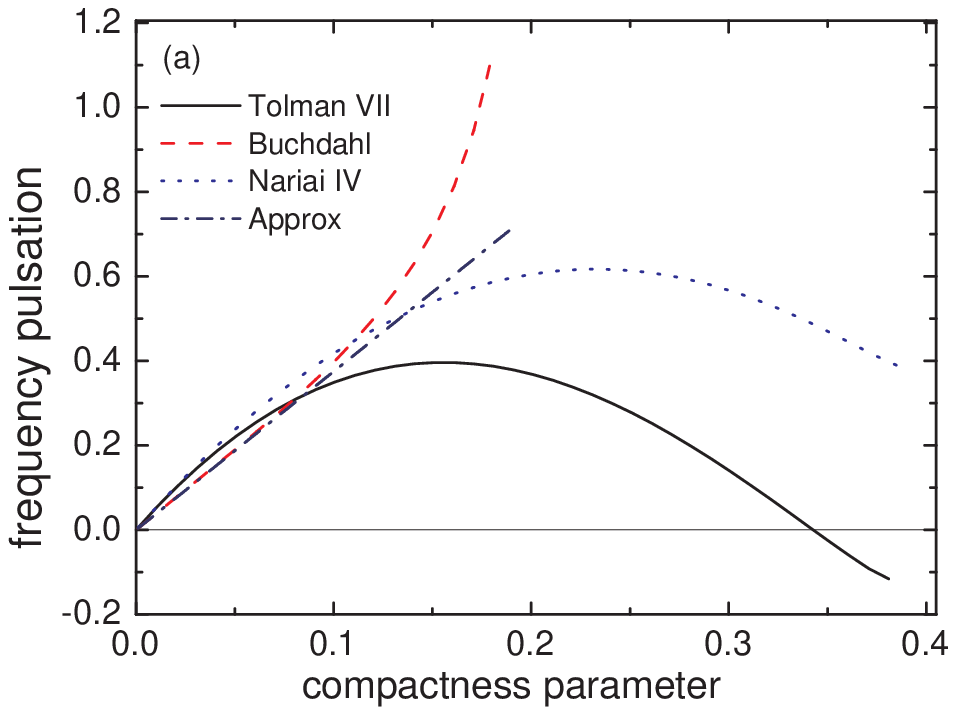}\
\includegraphics[height=8.5cm,width=8.5cm]{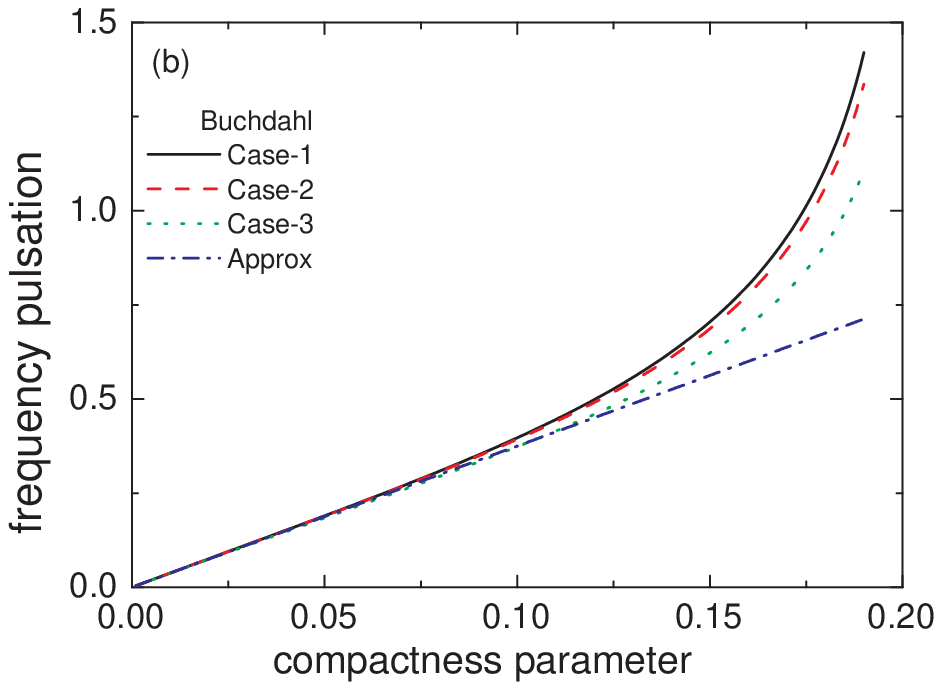}\
\caption{(a) The frequency pulsation $(\sigma R)^2$ as a function of
the compactness parameter $\beta=GM/Rc^2$ for the three analytical
solutions, Tolman VII, Buchdahl and Nariai IV, where the approximation (\ref{aprox-sigma}) is also included. (b) The frequency pulsation $(\sigma R)^2$ as a function of the
compactness parameter $\beta$ for the Buchdahl solution and for three different trial functions $\xi$. The case $1$ corresponds to the trial function  $\xi(r)=r e^{\nu/4}$, the case $2$ to  $\xi(r)=r e^{\nu/2}$ and case $3$ to $\xi(r)=b_1r\left(1+a_1r^2+2_2r^4+a_3r^6\right)e^{\nu/2}$. The approximation (\ref{aprox-sigma}) is also included for comparison.  } \label{Esnm-pnm}
\end{figure}

\begin{figure}
 \centering
\includegraphics[height=8.5cm,width=8.5cm]{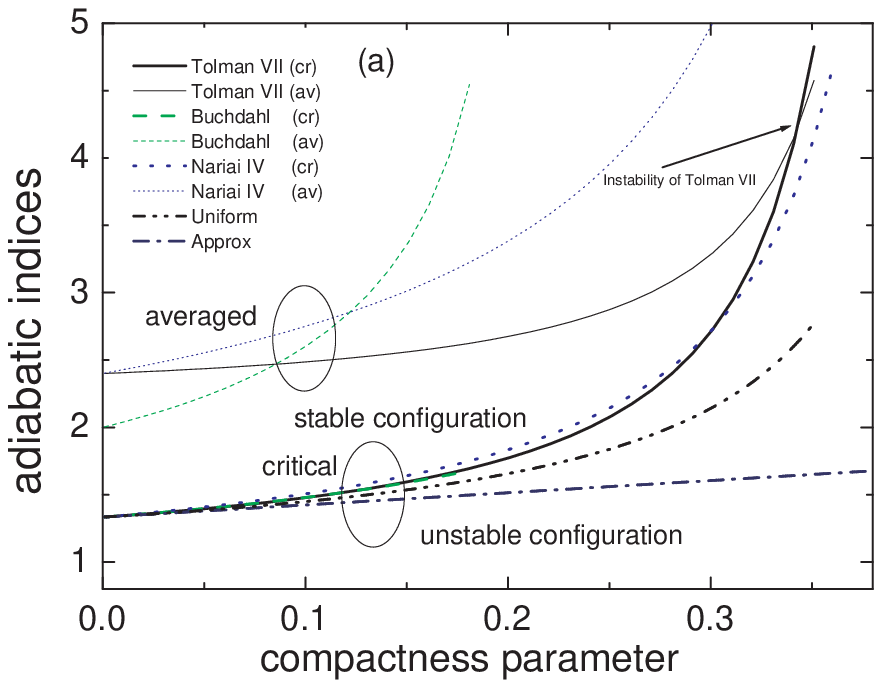}\
\includegraphics[height=8.5cm,width=8.5cm]{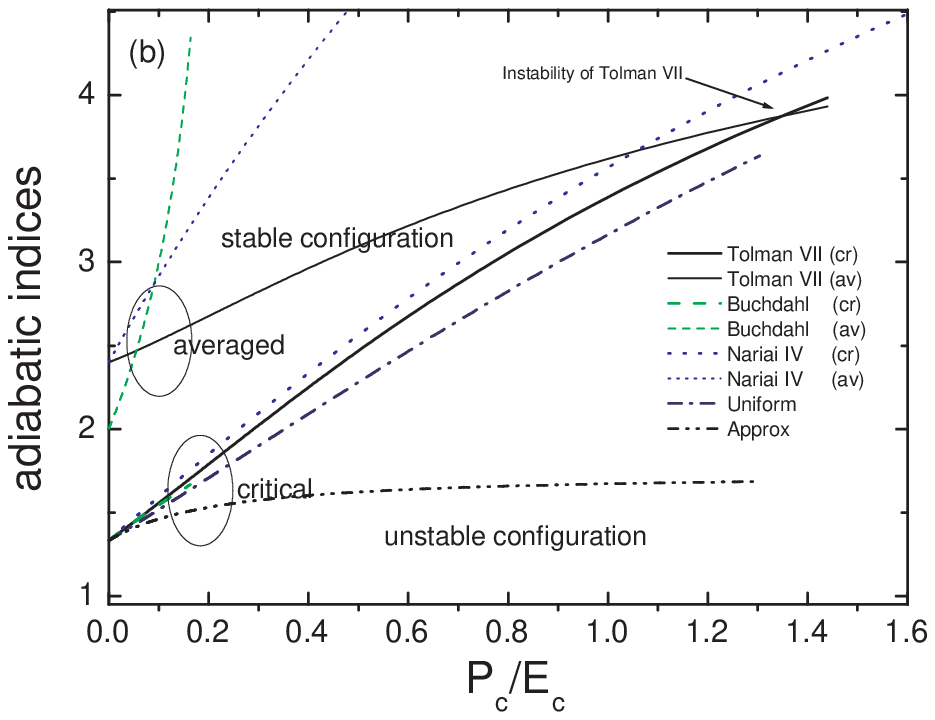}\
\caption{(a) The averaged adiabatic index $\langle \gamma \rangle$  and the critical adiabatic index $\gamma_{cr}$ as  a function of the
compactness parameter $\beta$  for the three analytical solutions. The results of the  uniform density solutions as well as the approximation~(\ref{gamma-Chns}) have been also included. The onset of instability for the Tolman VII solution is indicated.  (b) The averaged adiabatic index $\langle \gamma \rangle$
and the critical adiabatic index $\gamma_{cr}$ as  a function of
the ratio  $P_c/{\cal E}_c$  for the three analytical
solutions. The results of the uniform density solution as well as the approximation~(\ref{gcr-pcec-uni}) have been also included
for comparison.  } \label{Esnm-pnm}
\end{figure}

\begin{figure}
 \centering
\includegraphics[height=8.5cm,width=8.5cm]{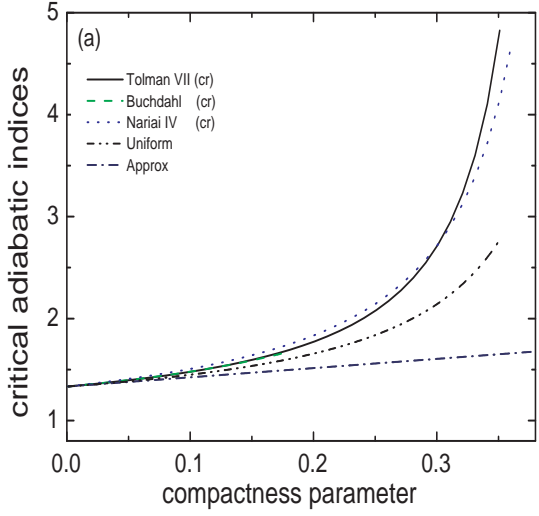}\
\includegraphics[height=8.5cm,width=8.5cm]{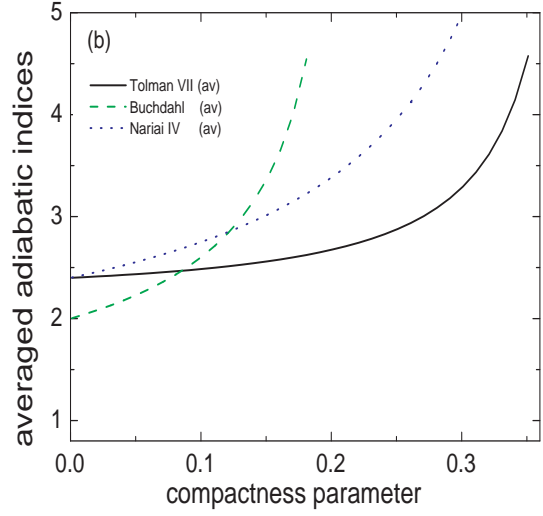}\
\includegraphics[height=8.5cm,width=8.5cm]{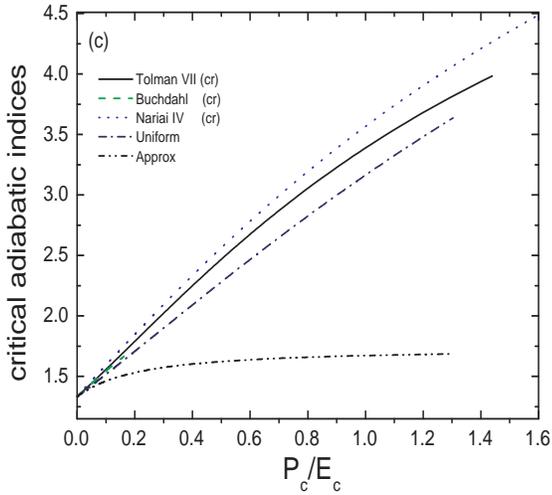}\
\includegraphics[height=8.5cm,width=8.5cm]{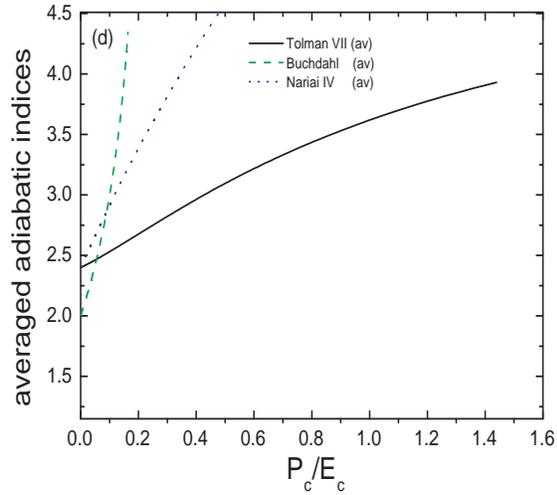}\
\caption{Same as in Fig.~5 where, for visual clarity, the dependence of the critical and averaged adiabatic indices  on  $\beta$ and $P_c/{\cal E}_c$, are plotted separately.   } \label{Esnm-pnm}
\end{figure}
\begin{figure}
 \centering
\includegraphics[height=8.1cm,width=9.5cm]{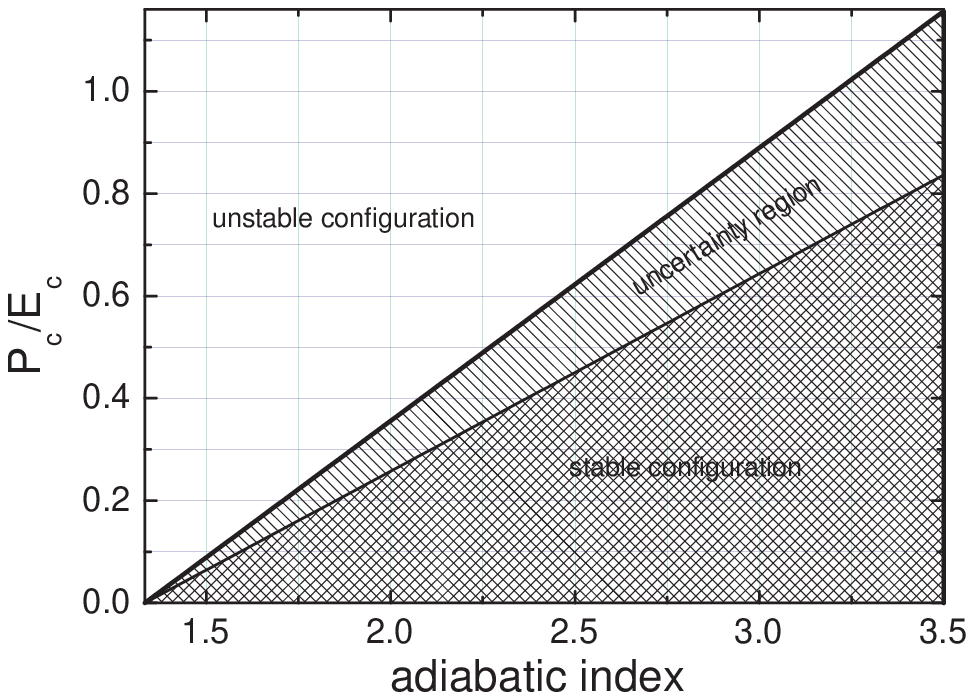}\
\caption{The stability (instability) window  of relativistic stars which defined by the analytical solutions employed in the present study. The upper curve corresponds to the equality of  (\ref{Eos-cons-2}) with  ${\cal K}=1.8734$ (introduced by the Schwarzschild constant-density interior solution) and the lower  one with  ${\cal K}=2.5904$ (introduced by the Nariai IV solution). The relative uncertainty corresponds to the intermediate region. For more details see text. }
\end{figure}


\begin{thebibliography}{99}
\bibitem{Chandrasekhar-64}S. Chandrasekhar, Astrophys. J., {\bf 140}, 417 (1964).
%
\bibitem{Chandrasekhar-64b}S. Chandrasekhar, Phys. Rev. Lett., {\bf 12}, 114 (1964).
%
\bibitem{Fowler-64}W.A. Fowler, Rev. Mod. Phys. {\bf 36}, 545 $\&$ 1104 (1964).
%
\bibitem{Fowler-66}W.A. Fowler, Astrophys. J., {\bf 144}, 180 (1966).
%
\bibitem{Bardeen-66}J.M. Bardeen, K.S. Thorne, and D.W. Meltzer, Astrophys. J., {\bf 145}, 505 (1966).
%
\bibitem{Zeldovich-65}Ya. B. Zel'dovichi and I.D. Novikov, Soviet Physics Uspekhi {\bf 84}, 763 (1965).
%
\bibitem{Tooper-65}R.F. Tooper, Astrophys. J., {\bf 142}, 1541 (1965).
%
\bibitem{Weinberg-72}S. Weinberg, {\it  Gravitational and Cosmology: Principle and Applications of the General Theory of Relativity}
(Wiley, New York, 1972).
%
\bibitem{Harrison-65} B.K. Harrison, K.S. Thorne, M. Wakano, J.A. Wheeler, {\it Gravitational Theory and Gravitational Collapse} (Chigago University  Press, 1965).
%
\bibitem{Zeldovich-78}Ya.B. Zeldovich and I.D. Novikov, {\it Relativistic Astrophysics, Vol.I} (University of Chigago Press, Chicago, 1978).
\bibitem{Shapiro-83}S.L. Shapiro and S.A. Teukolsky, {\it Black
Holes, White Dwarfs, and Neutron Stars} (John Wiley and Sons, New
York, 1983).
%
\bibitem{Glendenning-2000}N.K. Glendenning, {\it Compact Stars:
Nuclear Physics, Particle Physics, and General Relativity},
(Springer, Berlin, 2000).
%
\bibitem{Haensel-07}P. Haensel, A.Y. Potekhin, and D.G. Yakovlev,
{\it Neutron Stars 1: Equation of State and Structure}
(Springer-Verlag, New York, 2007).
%
\bibitem{Friedman-2013} J.L. Friedman and N. Stergioulas, {\it Rotating Relativistic Stars} (Cambridge University Press, 2013).
%

\bibitem{Tolman-39} R.C. Tolman, Phys. Rev. {\bf 55}, 364 (1939).
%
\bibitem{Buchdal-67}H.A. Buchdahl, Astrophys. J. {\bf 147}, 310 (1967).
%
\bibitem{Nariai-50}H. Nariai, Sci. Rep. Tohoku Univ. Ser. 1 {\bf 34}, 160 (1950).
\bibitem{Nariai-51}H. Nariai, Sci. Rep. Tohoku Univ. Ser. 1 {\bf 35}, 62 (1951).
\bibitem{Nariai-99}H. Nariai, Gen. Rel. and Grav. {\bf 31}, 951 (1999).
\bibitem{Lattimer-2001} J.M. Lattimer and M. Prakash, Astrophys. J., {\bf 550}, 426 (2001).
%
\bibitem{Lattimer-05}J.M. Lattimer, Neutron Stars, lectures delivered at the 33rd Summer Institute on Particle Physics, SSI 2005 (unpablished).
%
\bibitem{Lattimer-05a}J.M. Lattimer and M. Prakash, Phys. Rev. Lett., {\bf 94}, 111101 (2005).
%
\bibitem{Postnikov-010} S. Postnikov, M. Prakash, and J.M. Lattimer, Phys. Rev {\bf D} 82, 024016 (2010).
%
\bibitem{Raghoonundun-15}A.M. Raghoonundun and D.W. Hobill, Phys. Rev {\bf D}, 92, 124005 (2015).
%
\bibitem{Moustakidis-016}M.C. Papazoglou and C.C. Moustakidis, Astrophys. Space Sci. {\bf 361}, 98 (2016).
%
\bibitem{Negi-1999}P.S. Negi and M.C. Durgapal, Gen. Relativ. Gravitation {\bf 31}, 13 (1999).
%
\bibitem{Negi-2001}P.S. Negi and M.C. Durgapal, Astrophys. Space Sci. {\bf 275}, 185 (2001).
%
\bibitem{Schutz-85}B. F. Schutz, {\it  A First Course in General Relativity}, (Cambridge University Press, Cambridge, 1985).
%
%
\bibitem{Lattimer-2000}J. M. Lattimer and M. Prakash, Phys. Rep. {\bf 333-334}, 121 (2000).
%
\bibitem{Lattimer-07b} J. M. Lattimer and M. Prakash, Phys. Rep. {\bf 442}, 109 (2007).
%
\bibitem{Lattimer-10b}J.M. Lattimer, New Astr. Rev. {\bf 54}, 101 (2010).
%
\bibitem{Knutsen-88}H. Knutsen, Gen. Relativ. Gravitation {\bf 20}, 317 (1988).
%
\bibitem{Negi-2007}P.S. Negi, Gen. Relativ. Gravitation {\bf 39}, 529 (2007).
%
\bibitem{Bludman-73a} S.A. Bludman, Astrophys. J. {\bf 183}, 637 (1973).
%
\bibitem{Bludman-73b} S.A. Bludman, Astrophys. J. {\bf 183}, 649 (1973).
%
\bibitem{Ipser-70} J.R. Ipser, Astrophys. Space Sci. {\bf 7}, 361 (1970).
%
\bibitem{Oppenheimer-39}J.R. Oppenheimer and G.M. Volkoff, Phys.Rev. {\bf 55}, 374 (1939).
%
\bibitem{Kramer-1980}D. Kramer, H. Stephani, M.A. MacCallum, and E. Hertl, {\it Exact Solutions of Einstein's Field Equations}, (Deutsche Verlag der Wissenschaften, Berli/Cambridge University Press, Cambridge, 1980).
%
\bibitem{Delgaty-1998}M.S.R. Delgaty and K. Lake, Comp. Phys. Commun. {\bf 115}, 395 (1998).
%
\bibitem{Lake-2003}K. Lake, Phys. Rev. D {\bf 67}, 104015 (2003).
%
%
\bibitem{Merafina-89}M. Merafina and R. Ruffini, Astron. Astrophys.  {\bf 221}, 4 (1989).
%
 \bibitem{Herrera-89}L. Herrera, G. Le Denmat, and N.O. Santos, MNRS {\bf 237}, 257 (1989).
%
\bibitem{Herrera-94} R. Chan, L. Herrera, and  N.O. Santos, MNRS {\bf 267}, 637 (1994).
%
\bibitem{Yousaf-2015} M. Sharif and Z. Yousaf, Astrophys. Space Sci. {\bf 355}, 317 (2015).
%
\bibitem{Yousaf-2016} Z. Yousaf and M.Z. Bhatti, Eur. Phys. J. C {\bf 76}, 267 (2016).
%
%
\bibitem{Buchdal-59}H.A. Buchdahl, Phys. Rev. 116, 1027 (1959).
%
%
\bibitem{Mak-2013} M.K. Mak and T. Harko, Eur. Phys. J. C {\bf 73}, 2585 (2013).
%
\bibitem{Douchin-01} F. Douchin and P. Haensel, A $\&$ A, {\bf 380}, 151 (2001).
%






\end{thebibliography}
\end{document}